\documentclass{article}
\usepackage{authblk}

\usepackage{amsfonts, amsmath, amssymb}
\usepackage{graphicx,color}
\usepackage{listings}
\usepackage{algorithm}
\usepackage{algpseudocode}
\usepackage{cancel} 
\usepackage{wrapfig}
\usepackage{color}
 \usepackage{tikz}
\usepackage{url}

 \usetikzlibrary{arrows,automata, positioning}
 \tikzset{help lines/.style={step=1cm,black,very thin},
   nodedef/.style={scale=0.45},
 	  mybound/.style={loosely dashed, thick},
 	  axes/.style={->,>=triangle 60,semithick},
 	  textcomment/.style={scale=1},
 	  LUabstraction/.style={thick},
 	  Mabstraction/.style={thick, dotted},
 	  ta/.style={node distance=3cm,auto,->, >=stealth',
             scale=0.5, transform shape}
 }

\newcommand{\LLL}{\mathfrak{L}}

\newcommand{\ta}{\mathrm{TA}}
\newcommand{\nta}{\mathrm{NTA}}
\newcommand{\ntaeps}{\mathrm{eNTA}}

\newcommand{\Actions}{\Sigma}

\newcommand{\eActions}{\Sigma_{\epsilon}}

\newcommand{\Locs}{\mathcal{Q}}
\newcommand{\Clocks}{\mathcal{X}}
\newcommand{\ResetClocks}{\mathcal{X}_{rst}}
\newcommand{\Invariants}{\mathcal{I}}
\newcommand{\clk}{x}
\newcommand{\Qinit}{\mathcal{Q}_{init}}
\newcommand{\Qacc}{\mathcal{Q}_{accept}}

\newcommand{\Trans}{\mathcal{T}}
\newcommand{\Guards}{\mathcal{G}}

\newcommand{\ClockVal}{\mathcal{V}}
\newcommand{\zerov}{\textbf{0}}

\newcommand{\Naturals}{\mathbb{N}}

\newcommand{\PReals}{\mathbb{R}_{\geq 0}}

\newcommand{\powerset}[1]{\mathcal P \left({#1}\right) }

\newtheorem{theorem}{Theorem}[section]

\newtheorem{definition}{Definition}[section] 
\newtheorem{example}{Example}[section]

\title{Bounded Determinization of Timed Automata with Silent Transitions}
\author[1]{Florian Lorber} 
\author[2]{Amnon Rosenmann}
\author[2]{Dejan Ni\v{c}kovi\'{c}} 
\author[1]{Bernhard K. Aichernig}
\affil[1]{Institute for Software Technology\\ Graz University of Technology, Austria}
\affil[2]{AIT Austrian Institute of Technology GmbH \\Vienna, Austria}

\date{}
\begin{document}
\maketitle
\begin{abstract}
Deterministic timed automata are strictly less expressive than their
non-deterministic counterparts, which are again less expressive than
those with silent transitions. As a consequence, timed
automata are in general non-determinizable. This is unfortunate since
deterministic automata play a major role in model-based testing,
observability and implementability. However, by bounding the length of
the traces in the automaton, effective determinization becomes
possible. 
We propose a novel procedure for bounded
determinization of timed automata. The procedure unfolds the automata
to bounded trees, removes all silent transitions and determinizes via
disjunction of guards.  The proposed algorithms are optimized to the
bounded setting and thus are more efficient and can handle a larger
class of timed automata than the general algorithms. The approach is
implemented in a prototype tool and evaluated on several examples. To
our best knowledge, this is the first implementation of this type of
procedure for timed automata.   
\end{abstract}

\section{Introduction}
\label{sec:intro}

The design of modern embedded systems often involves the integration of interacting components $I_1$ and $I_2$ that realize some requested behavior.
In early stages of the design, $I_{1}$ and $I_{2}$ are high-level and partial models that allow considerable implementation freedom to the designer.
In practice, this freedom is reflected in the non-deterministic choices that are intended to be resolved during subsequent design refinement steps.
In addition, the composition of two components involves their synchronization on some shared actions.
Typically, the actions over which the two components interact
are {\em hidden} and become unobservable to the user. 
It follows that the overall specification $I = I_{1}~||~I_{2}$ can be a {\em non-deterministic partially observable} model.
%
%
However, for many problems such as model-based testing, observability, implementability and 
language inclusion checking, it is desirable and in certain cases necessary 
to work with the deterministic model.

Many embedded systems must meet strict real-time requirements.
Timed automata (TA)~\cite{ta} are a formal modeling language that enables specification of complex real-time systems.
In contrast to the classical automata theory,
deterministic TA (DTA) are strictly less expressive than the fully observable non-deterministic TA (NTA)~\cite{ta,tripakis,Finkel06}, 
whereas the latter are strictly less expressive than TA with silent transitions (eNTA)~\cite{ta-eps}.
This strict hierarchy of TA with respect to determinism and observability has an important direct consequence - NTA are not determinizable in general. 
%
In addition, due to their complexity, it is rarely the case that exhaustive verification methods are used during the design of modern embedded systems.
Lighter and incomplete methods, such as model-based testing~\cite{mbt} and bounded model checking~\cite{bmc} are used in practice in order to gain confidence in the design-under-test and effectively catch bugs.

In this paper, we propose a procedure for {\em bounded determinization} of $\ntaeps$.
Given an arbitrary {\em strongly responsive}\footnote{In model-based testing, strong responsiveness is the requirement that there are no silent loops, otherwise
the tester cannot distinguish between deadlocks and livelocks.}
$\ntaeps$ $A$
and a bound $k$, our algorithm computes a DTA $D(A)$ in the form of a timed tree, such that every timed trace consisting of at most $k$ observable actions is a trace in $A$ if and only if it is a trace in $D(A)$.
It provides the basis for effectively implementing bounded refinement checking and test case generation procedures.

Our concrete motivation behind determinizing the model was induced by our previous model-based testing approach~\cite{DBLP:conf/tap/AichernigLN13}. This approach uses fault-based techniques for the test generation and needs to perform language-inclusion between correct and faulty timed automata models. The language inclusion is implemented via SMT-solving  and relies on deterministic models. Thus, the determinization enables the processing of a wider class of models and the restriction to bounded traces does not pose a problem, as testing only considers finite traces. 


The proposed algorithms are performed in three steps: (1) we unfold the original automaton into a finite tree and rename the clocks in a way that only needs one clock reset per transition, (2) we remove the silent transitions from the tree, (3) we determinize it.
Our determinization procedure results in a TA description which includes diagonal~\cite{bouyer2} and disjunctive constraints.
Although non-standard, this representation is practical and optimized for the bounded setting -- 
it avoids costly transformation of the TA into its standard form 
and exploits efficient heuristics in SMT solvers that can directly deal with this type of constraints. 
In addition, our focus on bounded determinization allows us to consider models, such as TA with loops containing both observable and silent transitions with reset, that could not be determinized otherwise.
We implemented the procedure in a prototype tool and evaluated it on several examples. 
To our best knowledge, this is the first implementation of this type of procedure for timed automata.

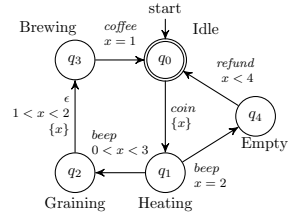
\begin{wrapfigure}{RB}{40mm}
\vspace{-18pt}
\begin{minipage}{0.3\linewidth}
\begin{tikzpicture}[ta, scale=1.2]
  \node[state] [state,initial above,accepting] (p1) at (0,0) {$q_0$}; 
  \node[state]  (p2) at (0,-2.5) {$q_1$}; 
  \node[state]  (p3) at (2,-1.25) {$q_4$}; 
  \node[state]  (p4) at (-2,-2.5) {$q_2$}; 
  \node[state]  (p5) at (-2,0) {$q_3$}; 
  \node at (0,-3.2) {Heating};
  \node at (0.9,0.7) {Idle};
  \node at (2.2,-1.87) {Empty};
  \node at (-2,-3.2) {Graining};
  \node at (-2.6,0.7) {Brewing};

  \path [every node/.style={font=\footnotesize}]
      (p1) edge node[right] [align=left] {$coin$ \\ 
        $\{x\}$} (p2)
      (p2) edge node [below=8pt] [align=left]  {$beep$ \\
        $x=2$} (p3)
      (p2) edge node[above=8pt][align=left]  {$beep$ \\ $0 < x < 3$} (p4) 
      (p4) edge node[left][align=left] [align=right] {$\epsilon$ \\ $1 < x < 2$ \\$\{x\}$} (p5)
      (p3) edge node[above right][align=right]  {{\it refund} \\ $x < 4$} (p1) 
      (p5) edge node[above= 6pt][align=right]  {{\it coffee} \\ $x=1$} (p1) 
;
\end{tikzpicture}
\end{minipage}
\label{fig:coffee1}
\vspace{6pt}
\caption{Running example}
\end{wrapfigure} 
\noindent \textbf{Running example.}
The different steps of the algorithms will be illustrated on a running example of a coffe-machine shown in Figure \ref{fig:coffee1}. After inserting a \textit{coin}, the system heats up for zero to three seconds, followed by a \textit{beep}-tone indicating its readyness. Alternatively, if there is no coffee or water left, the \textit{beep} might occur after exactly two seconds, indicating that the \textit{refunding} process has started and the coin will be returned within four seconds. Heating up and graining the coffee together may only take between one and two seconds. Then the brewing process starts and finally the machine releases the \textit{coffee} after one second of brewing. There is no observable signal indicating the transition from graining to brewing, thus this transition is silent.


The rest of the paper is structured as follows:
First, we give the basic definitions and notation of TA with silent transitions (Section~\ref{sec:ta}).
Then,  we illustrate the first step of our procedure, the bounded-unfolding of the automaton and the renaming of clocks (Section~\ref{sec:k-bounded-unfolding}).  This is followed by the second step, the removal of silent transitions (Section~\ref{sec:rem_silent_trans}) and the final step, our determinization approach (Section~\ref{sec:bdtawst}).
Section~\ref{sec:complexity} summarizes the complexity of the different steps.
In Section~\ref{sec:exp_res} we evaluate our prototype implementation
and in Section ~\ref{sec:related_work} we address related work.
Finally, in Section~\ref{sec:conclusion} we conclude our work.
Complete proofs of the propositions and theorems can be found in the appendices.

\section{Timed Automata with Silent Transitions}
\label{sec:ta}


A timed automaton is an abstract model aiming at capturing the real-time behaviour of systems.
It is a finite automaton extended with a set of clocks defined over
$\PReals$, the set of non-negative real numbers.
We may represent the timed automaton by a graph whose nodes are called {\em locations}, which are defined through a set of upper bounds put on the clock values.
These bounds are restricted to non-negative integer values.
While being at a location, all clocks progress at the same rate.
The edges of the graph are called  {\em transitions}.
Each transition may be subject to constraints, called {\em guards}, put on clock values in the form of integer inequalities.
At each such transition an {\em action} occurs and some of the clocks may be reset.
The actions take values in some finite domain denoted by $\Actions$.
Here we are dealing with the class of timed automata with an extended set of actions including also {\em silent actions}, denoted by $\epsilon$.
hese are internal actions that are {\em non-observable} from the outside, 
and we distinguish them from the actions that are not silent and called {\em observable} actions.
We call a TA without silent transitions {\em fully-observable}.

Let $\Clocks$ be a finite set of {\em clock} variables.
A clock {\em valuation} $v(\clk)$ is a function $v:\Clocks \to \PReals$ assigning a real value to every clock $\clk \in \Clocks$.
We denote by $\ClockVal$ the set of all clock valuations and by $\zerov$ the valuation assigning $0$ to every clock.
For a valuation $v$ and $d \in \PReals$ we define $v+d$ to be the valuation $(v+d)(\clk) = v(\clk)+d$ for all $ \clk \in \Clocks$.
For a subset $\ResetClocks$ of $\Clocks$, we denote by $v[\ResetClocks]$ the valuation such that for every $\clk \in\ResetClocks$,
$v[\ResetClocks](\clk) = 0$ and for every $\clk \in \Clocks \setminus \ResetClocks$, $v[\ResetClocks](\clk) = v(\clk)$.
A {\em clock constraint} $\varphi$ is a conjunction of predicates of the form $\clk \sim n$, where $\clk \in \Clocks$, $n \in \Naturals$ and $\sim \ \in \{ <, \leq, =, \geq, > \}$.
Given a clock valuation $v$, we write $v \models \varphi$ when $v$ satisfies $\varphi$.
We give now a formal definition of (non-deterministic) timed automata with silent transitions. 
\begin{definition}[$\ntaeps$]
\label{def:ta}
A (non-deterministic) timed automaton with silent transitions $A$ is a tuple $(\Locs, q_{init}, \eActions, \Clocks, \Invariants,
\Guards,\Trans,  \Qacc)$, where
$\Locs$ is a finite set of {\em locations} and $q_{init} \in \Locs$ is the {\em initial} location;
$\eActions = \Sigma \cup \{\epsilon\}$ is a finite set of actions, where $\Actions$ are the observable actions and $\epsilon$ represents
a silent action, that is a non-observable internal action;
$\Clocks$ is a finite set of {\em clock} variables;
$\Invariants: L \rightarrow LI$ is a mapping from locations to {\em
  location invariants}, where each location invariant $li \in LI$ is a
conjunction of constraints of the form $true$, $\clk < n$ or $\clk \leq n$, with $\clk \in \Clocks$ and $n \in \Naturals$;
$\Guards$ is a set of {\em transition guards}, where each guard is a conjunction of constraints of the form $\clk \sim n$, where $\clk \in \Clocks$, $\sim \ \in \{<,\leq, =,\geq, >\}$ and $n \in \Naturals$;
$\Trans \subseteq \Locs \times \eActions \times \Guards \times \powerset{\Clocks} \times \Locs$ is a finite set of {\em transitions} of the form $(q, \alpha, g, \ResetClocks, q')$, where
%
$q,q' \in \Locs$ are the {\em source} and the {\em target} locations;
$\alpha \in \eActions$ is the transition {\em action};
$g \in  \Guards$ is the transition guard;
$\ResetClocks \subseteq \Clocks$ is the subset of clocks to be {\em reset};
$\Qacc \subseteq \Locs$ is the subset of {\em accepting} locations.
\end{definition}
%

\begin{example}
For the eNTA illustrated in Figure~\ref{fig:coffee1} we have $\Locs = \{q_0, \dots, q_4\}$, $q_{init} = q_0$, $\eActions =
\{\epsilon, \mathit{coin}, \mathit{beep}, \mathit{refund}, \mathit{coffee}\}$, $\Clocks = \{x\}$,
$\Invariants(q_i) = true | q_i \in \Locs$, $\Guards = \{0<x<3, x=2, x<4, 1<x<2, x=1\}$,
$\Qacc = \{q_0\}$. $\Trans$ is the set containing all transitions,
e.g. the transition from $q_2$ to $q_3$, with $\alpha = \epsilon$ (thus,
it is a silent transition), $g = 1<x<2$ and $\ResetClocks = \{x\}$.
\end{example}

The {\em semantics} of an $\ntaeps$ $A$ is given by the {\em timed transition system}
$[[A]] = (S, s_{init}, \PReals, \eActions, T, S_{accept})$, where
$S = \{(q,v) \in \Locs \times \ClockVal ~|~ v \models \Invariants(q) \}$;
$ s_{init} = (q_{init},\zerov)$;
$T \subseteq S \times (\eActions \cup \PReals) \times S$ is the transition relation consisting of  {\em timed} and {\em discrete} transitions such that:
\emph{Timed transitions (delay):} $((q,v), d, (q, v+d)) \in T$, where $d \in \PReals$, if $v+d \models \Invariants(q)$;
\emph{Discrete transitions (jump):} $((q,v), \alpha, (q',v')) \in T$, where $\alpha \in \Actions$, if there exists a transition
$(q, \alpha, g, \ResetClocks, q')$ in $\Trans$, such that: (1) $v \models g$; (2) $v' = v[\ResetClocks]$ and (3) $v' \models \Invariants(q')$;
$S_{accept} \subseteq S$ such that  $(q,v)\in S_{accept}$ if and only if $q \in \Qacc$.

A {\em finite well-behaving run} $\rho$ of an $\ntaeps$ $A$ is a finite sequence of alternating timed and discrete transitions, that ends with an observable action, of the form
$(q_{0}, v_{0}) \xrightarrow{d_{1}} (q_{0}, v_{0} + d_{1}) \xrightarrow{\tau_{1}} (q_{1}, v_{1}) \xrightarrow{d_{2}} \cdots 
\xrightarrow{d_{n}} (q_{n-1}, v_{n-1} + d_{n}) \xrightarrow{\tau_{n}} (q_{n}, v_{n})$,
where $q_{0} = q_{init}$, $v_{0} = \zerov$, $\tau_{i} = (q_{i-1}, \alpha_{i}, g_{i},  \Clocks_{rst(i)}, q_{i}) \in\Trans$ and $\alpha_i \in \Actions$.
In this paper we consider only finite and well-behaving runs.
A run $\rho$ is {\em accepting} if the last location $q_n$ is accepting.
The run $\rho$ of $A$ induces the {\em timed trace}
$\sigma = (t_{1}, \alpha_{1}), (t_{2}, \alpha_{2}), \ldots, (t_{n}, \alpha_{n})$
defined over $\eActions$, where $t_{i} = \Sigma_{j=1}^{i} d_i$. 
From the latter we can extract the {\em observable timed trace}, which is obtained by removing from $\sigma$ all the pairs containing silent actions 
while taking into account the passage of time.
A TA is called {\em deterministic} if it does not contain silent transitions
and whenever two timed traces are the same then they are induced by the same run.
Otherwise, the TA is {\em non-deterministic}.
The {\em language} accepted by an $\ntaeps$ $A$, denoted $\LLL(A)$, is the set of observable timed traces induced by all accepting runs of $A$.
%
Note, that the restriction to well-behaving runs is compatible with the definition of the language of the automaton, where silent actions that occur after the last observable action on a finite run are ignored.
As a consequence, a location with in-going edges consisting of only silent transitions cannot be an accepting location.

%
%
%
\section{$k$-Bounded Unfolding of Timed Automata}
\label{sec:k-bounded-unfolding}

Given an $\ntaeps$ $A$ which is strongly responsive, its $k$-prefix language $\LLL_k(A) \subseteq \LLL(A)$ is the set of observable timed traces induced by all accepting runs of $A$ 
\begin{wrapfigure}{LT}{62mm}
\begin{minipage}{0.49\linewidth}
\begin{tikzpicture}[ta, scale=1.2]
  \node[state] [state,initial,accepting] (p1) at (0,0) {$q_0$}; 
  \node[state]  (p2) at (0,-2) {$q_1$}; 
  \node[state]  (p3) at (1,-4) {$q_4$}; 
  \node[state]  (p4) at (-1,-4) {$q_2$}; 
  \node[state]  (p5) at (-1,-6) {$q_3$}; 
  \node[state,accepting]  (p6) at (1,-6) {$q_5$}; 
  \node[state,accepting]  (p7) at (-1,-8) {$q_6$}; 
  \node at (0,-8) {(a)};
  \path [every node/.style={font=\footnotesize}]
      (p1) edge node[left=15pt] [align=right] {$coin$ \\ 
        $\{x\}$} (p2)
      (p2) edge node [above=10pt][right=8pt] [align=left]  {$beep$ \\
        $x=2$} (p3)
      (p2) edge node[above left = 1pt and 5pt][align=right]  {$beep$ \\ $0 < x < 3$} (p4) 
      (p4) edge node[left][align=left]  {$\epsilon$ \\ $1 < x < 2$ \\$\{x\}$} (p5)
      (p3) edge node[left = 7pt][align=right]  {{\it refund} \\ $x < 4$} (p6) 
      (p5) edge node[left = 7pt][align=right]  {{\it coffee} \\ $x=1$} (p7) ;
\end{tikzpicture}
\end{minipage}
\begin{minipage}{0.49\linewidth}
\begin{tikzpicture}[ta, scale=1.2]
  \node[state] [state,initial,accepting] (p1) at (0,0) {$q_0$}; 
  \node[state]  (p2) at (0,-2) {$q_1$}; 
  \node[state]  (p3) at (1,-4) {$q_4$}; 
  \node[state]  (p4) at (-1,-4) {$q_2$}; 
  \node[state]  (p5) at (-1,-6) {$q_3$}; 
  \node[state,accepting]  (p6) at (1,-6) {$q_5$}; 
  \node[state,accepting]  (p7) at (-1,-8) {$q_6$}; 
  \node at (0,-8) {(b)};
  \path [every node/.style={font=\footnotesize}]
      (p1) edge node[left=15pt] [align=right] {$coin$ \\ 
        $\{x_1\}$} (p2)
      (p2) edge node [above=10pt][right=8pt] [align=left]  {$beep$ \\
        $x_1=2$ \\ $\{x_2\}$} (p3)
      (p2) edge node[above left = 1pt and 5pt][align=right]  {$beep$ \\ $0 < x_1 < 3$ \\ $\{x_2\}$} (p4) 
      (p4) edge node[left][align=left]  { $\epsilon$ \\ $1 < x_1 < 2$ \\$\{x_{2,0}\}$} (p5)
      (p3) edge node[left = 7pt][align=right]  {{\it refund} \\ $x_1 < 4$ \\$\{x_3\}$} (p6) 
      (p5) edge node[left = 7pt][align=right]  {{\it coffee} \\ $x_{2,0}=1$ \\ $\{x_3\}$} (p7) ;
\end{tikzpicture}
\end{minipage}

\vspace{4pt}
\caption{Unfolding and clock renaming}
\label{fig:coffe2}
\vspace{10pt}
\end{wrapfigure}
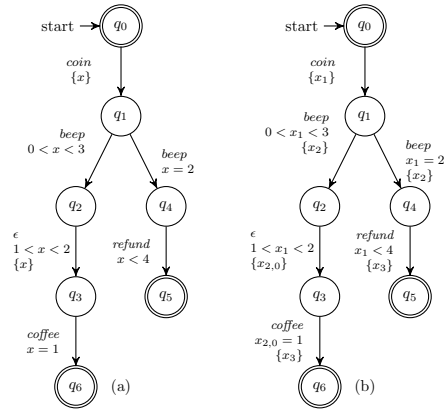

\noindent which are of observable length bounded by $k$.  That is,
\begin{equation}
\LLL_k(A) = \{ w \in \LLL(A) \, | \, | w | \leq k \}.
\end{equation}
By unfolding $A$ and cutting it at observable level $k$, the resulting TA, $U_k(A)$, satisfies
\begin{equation}
\LLL(U_k(A)) = \LLL_k(A).
\end{equation}

$U_k(A)$ is in the form of a finite tree, where each path that starts at the root ends after at most $k$ observable transitions, and we may also further cut $A$ by requiring that all leaves are accepting locations.
Note, that if we reach in $U_k(A)$ a copy of an accepting location $q$
of $A$ by a silent transition then it will not be marked as an
accepting location (but another copy might be marked as an accepting
location if reached by an observable transition).

Figure \ref{fig:coffe2}(a) shows the unfolding of the coffee-machine up
to observable depth three. The left branch is longer than the right,
as it contains a silent transition.
\subsection{Renaming the Clocks}
Every unfolded timed automaton can be expressed by an
equivalent timed automaton that resets at most one clock per
transition. This known normal form  \cite{bbbb} crucially simplifies the next stages of our algorithm, where we
do not need to bother with multiple clock resets in one transition. 
The basic idea is to substitute the clocks from the original automaton
by new clocks, where multiple old clocks reset at the same transition are
replaced by the the same new clock, as they measure the same time
until they are reset again. 
The substitution of the clocks works straight forward: At each path from the root, at the $i$-th observable transition, a new
clock $x_i$ is introduced and reset, and if this transition is
followed by $l > 0$  silent transitions then new clocks $x_{i,0},
\ldots, x_{i,l-1}$ are introduced and reset.  A clock $x$ that occurs in a guard is substituted by the new clock that
was introduced in the transition where the last reset of $x$
happened, or by $x_0$ if it was never reset.
Let $\tau_i$ and $\tau_j$ be two transitions on the same path
in the original automata at observable depth $i, j$, s.t. $i<j$. 
Furthermore, a clock $x$ appearing in the
guard of $\tau_j$, is reset before in $\tau_i$, but is not reset on any
transition in between $\tau_i$ and $\tau_j$. Then, $x_i$ is introduced and
reset at $\tau_i$ and the original clock variable $x$ is substituted by
$x_i$ in the guard of $\tau_j$.
Figure \ref{fig:coffe2}(b) illustrates the clock renaming applied to
the coffee machine. In the guards of the two \textit{beep}-transitions
starting at $q_1$, $x$ is replaced by $x_1$, since the last reset of $x$ in
the original automata was at depth one, while in the
\textit{coffee}-transition from $q_3$ it is replaced by $x_{2,0}$, as
$x$ was reset in the first silent transition after depth two.

\section{Removing the Silent Transitions}
\label{sec:rem_silent_trans}

\begin{figure}[th]
\centering
\scalebox{0.28}{ 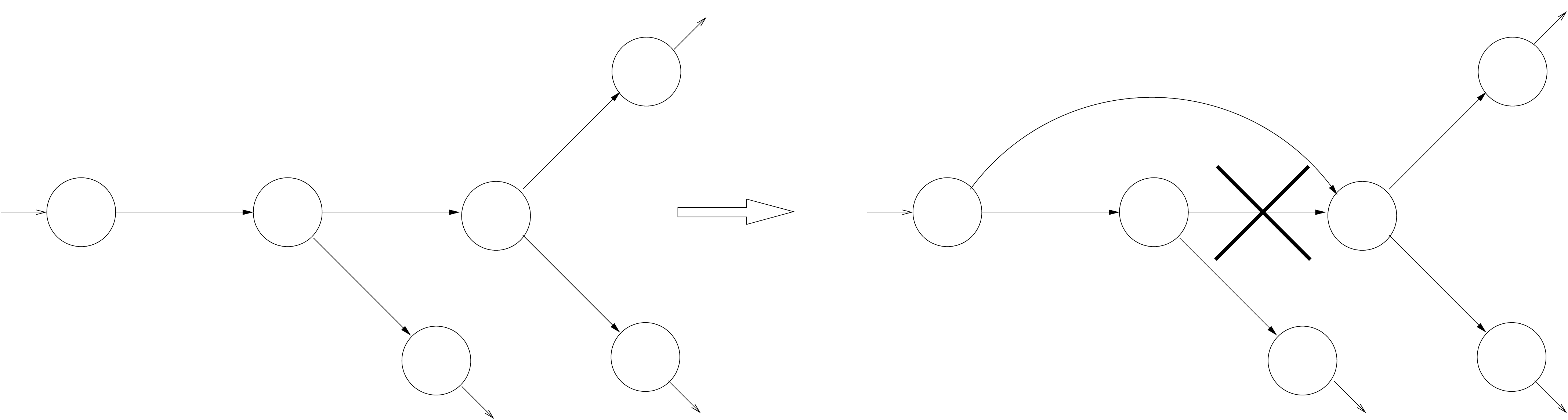 }
\caption{Bypassing the silent transition}
\label{fig:bypass1}
\end{figure}

In this section we give an algorithm that removes the silent
transitions from the $\ntaeps$ $A$, which is in the form of a finite
tree with renamed clocks. Thus, at each level $i$ there will
be a single clock $x_i$ reset on all transitions of that level.
Algorithm~\ref{alg:rst} shows the workflow and
Figure \ref{fig:bypass1} illustrates the general idea.
\begin{figure}[hb]
\vspace{-15pt}
\begin{minipage}[t]{1.0 \linewidth}
\begin{algorithm}[H]
\begin{algorithmic}[1]
\Require $A \in\ntaeps_k$ in the form of a tree of observable depth
$k$ with renamed clocks
\Ensure $O(A) \in \nta_k$, such that $\LLL(O(A)) = \LLL(A)$
\While{there are silent transitions}
	\State \textsc{Find} first (from root) silent transition $\tau_{s,0}$ from $q_s$ to $q_{s,0}$ 
	\State \textsc{Set} lower bound to the silent transition
	\State \textsc{Create} bypass transition with enabling guard
	\State \textsc{Augment} transitions from $q_{s,0}$ with taken guard
	\State \textsc{Update} guards on paths from $q_{s,0}$
	\State \textsc{Remove} $\tau_{s,0}$ 
\EndWhile
\end{algorithmic}
\caption{Removing the Silent Transitions}
\label{alg:rst}
\end{algorithm}
\end{minipage}
\end{figure}

We remove the silent transitions one at a time, where at each iteration we remove the first occurrence of a silent transition on some path from the root, until no silent transitions are left (e.g. we can pick a path and move one-by-one all its silent transitions, then move to another path, and so on).
So, let $\tau_{s,0}$ be such a first silent transition found by Line $2$
of the algorithm, leading from location $q_s$ to location $q_{s,0}$ with guard $g_{s,0}$ and reset of clock $x_{s,0}$.
Let $q_s$ be reached from location $q_{s-1}$ with an observable transition $\tau_s$ and with guard $g_s$.
The case where $q_s$ is the initial location is simpler, as it does not
require building a bypass transition. 
In order to remove the silent transition  $\tau_{s,0}$ after forming a transition
that bypasses it, several steps are carried out, that will be
explained in detail in the following subsections.
First, we set an auxilliary lower bound on the clock that is reset on
the silent transition by updating the guard (Line $3$). Then,
we create the bypass transition using an \emph{enabling guard}
$eg(\tau_{s,0})$ which represents the upper bound until when the silent transition
$\tau_{s,0}$ is enabled 
(Line $4$).
In Line $5$ we construct a \emph{taken guard} $tg(\tau_{s,0})$ that ensures that
the transitions from $q_{s,0}$ come after the necessary delay that is
forced by the silent transition. The taken guard is added to all transitions leaving
$q_{s,0}$.
Finally, in Lines $6$--$7$, we remove the silent transition $\tau_{s,0}$  and update all future guards referring to
the deleted clock $x_{s,0}$.
%

\begin{wrapfigure}{LT}{40mm}
\begin{minipage}{0.18\linewidth}
\begin{tikzpicture}[ta, scale=1.2]
  \node[state] [state,initial,accepting] (p1) at (0,0) {$q_0$}; 
  \node[state]  (p2) at (0,-2) {$q_1$}; 
  \node[state]  (p3) at (2.5,-4) {$q_4$}; 
  \node[state]  (p4) at (-2.5,-4) {$q_{3}$}; 
  \node[state]  (p5) at (0,-4) {$q_2$}; 
  \node[state,accepting]  (p6) at (2.5,-6) {$q_5$}; 
  \node[state,accepting]  (p7) at (-2.5,-6) {$q_6$}; 
  \path [every node/.style={font=\footnotesize}]
      (p1) edge node[right=6pt] [align=right] {$coin$ \\ 
        $\{x_1\}$} (p2)
      (p2) edge node [above right=8pt and 10pt] [align=left]  {$beep$ \\
        $x_1=2$ \\ $\{x_2\}$} (p3)
      (p2) edge node [above left =-2pt and -7pt] [align=left]  {$beep$ \\ $0 < x_1 < 3$\\
        \textcolor{red} {$\wedge$} \textcolor{red} {$x_1 < 2$} \\ $\{x_2\}$} (p4) 
      (p2) edge node[right=1pt][align=left]  {$beep$ \\ $0 < x_1 < 3$ \\$\{x_{2}\}$} (p5)
      (p3) edge node[left=7pt][align=right]  {{\it refund} \\ $x_1 < 4$ \\$\{x_3\}$} (p6) 
      (p4) edge node[right = 9pt][align=left]  {{\it coffee} \\ $2<x_1<3$ \\  \textcolor{red} {$\wedge$} \textcolor{red} {$1 < x_1$} \\ $\{x_3\}$} (p7) ;
\end{tikzpicture}
\end{minipage}
\vspace{5pt}
\caption{Fully observable non-deterministic TA}
\label{fig:coffee3}
\vspace{-10pt}
\end{wrapfigure}
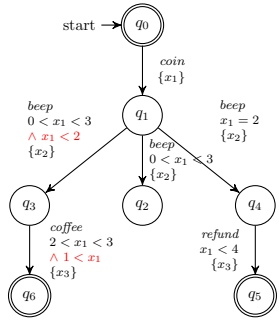

\subsubsection{Setting a Lower Bound to the Silent Transition.}
We set a lower bound to the silent transition by augmenting the guard
$g_{s,0}$ of $\tau_{s,0}$ to be $g'_{s,0} = g_{s,0} \wedge (0 \leq
x_s)$, where $x_s$ is the clock that is reset on the transition
$\tau_s$ that precedes the silent transition, thus ensuring that when
referring to the guard of the silent transition we do not refer to an
earlier time.
This additional constraint per definition always evaluates to $true$,
but it is used in the next step to compute the unary constraints of
the enabling guard. The guard of the silent transition in Figure
\ref{fig:coffe2} (b) after setting the lower bound is $1 < x_1 < 2
\wedge 0 \leq x_2$. 
%
\subsubsection{Creating a Bypass with the Enabling Guard.}
The enabling guard $eg(\tau_{s,0})$ guarantees that each clock's
constraint that was part of the silent transition 
 is satisfied at some non-negative delay and
that these constraints are satisfied simultaneously, thus at some
point during the bypass transition the silent transition would have
been enabled as well. 
We describe here how the enabling guards are defined for strict inequalities, as shown in the upper part of Table~\ref{table:enabling_guard}.
The other cases are dealt similarly, as seen in the table, and the constraint $x_i = n_i$ is treated as $n_i \leq x_i \leq n_i$.
For every pair of a lower bound constraint $m_i < x_i$ and an upper
bound constraint $x_j < n_j$, where $i \neq j$ and $x_i, x_j \neq
x_s$ ($x_s$ is the clock that is reset at $\tau_s$), that appear in
$g'_{s,0}$ we form the enabling guard binary constraint $x_j - x_i <
n_j - m_i$ as shown in the first line of Table~\ref{table:enabling_guard}.

The next two lines consider constraints that involve the clock $x_s$,
where $x_s$ will be removed as it is the clock that will be reset on
the bypass and is considered of value $0$. Note, that for each upper bound constraint $x_j < n_j$ we
use the lower bound constraint $0 \leq x_s$ that was added in the
previous step of the algorithm to compute the enabling guard unary
constraint $x_j < n_j$, which guarantees that at the time of the
bypass $x_j$ does not pass its upper bound constraint of the silent
transition. An example of such a unary constraint is
marked in red in the transition from $q_1$ to $q_3$ in Figure
\ref{fig:coffee3}. The silent transition in the original automaton could
not have been enabled if $x_1$ had already been higher than two after
the $beep$-transition, thus the bypass can also only be enabled while
$x_1$ is smaller than two. The running example does not contain any binary constraints.

To create the bypass, we split the paths through $q_s$ in the original automaton $A$ into two.
Those that do not take the silent transition $\tau_{s,0}$ continue as before from $q_{s-1}$ to $q_s$ and then to some location different from $q_{s,0}$.
The paths that went through $\tau_{s,0}$ are directed from $q_{s-1}$ to $q_{s,0}$ and then continue as before.
The bypass  $\tau'_s$ from $q_{s-1}$ to $q_{s,0}$ has the same
observable actions as those of $\tau_s$, the same new clock reset
$x_s$, and the guard $g'_s$ which is the guard $g_s$ of $\tau_s$
augmented with the {\em enabling guard} $eg(\tau_{s,0})$ (see
Figure~\ref{fig:bypass1}).  Figure~\ref{fig:coffee3} shows the removal
of the silent transition illustrated on the coffee-machine. The
transition from $q_1$ to $q_{3}$ is the bypass and the transition
from $q_1$ to $q_2$ is the original transition. Since the silent
transition was the only transition leaving $q_2$, $q_2$ does not
contain any outgoing transitions anymore, once the bypass is
generated. 
\begin{table}[t]
\centering
\begin{tabular}{ l | l | l }
\hline
{\bf Silent Trans. Constraints}  & {\bf Clock Reset} & {\bf Enabling Guard Constraint} \\
\hline
$(m_i < x_i) \wedge (x_j  < n_j)$ & $x_s$  & $x_j - x_i  < n_j - m_i$  \\
$(m_s < x_s) \wedge (x_j  < n_j)$ & $x_s$  & $x_j < n_j - m_s$  \\
$(m_i < x_i) \wedge (x_s  < n_s)$& $x_s$ &  $m_i - n_s < x_i$  \\
\hline
$(m_i \leq x_i) \wedge (x_j  < n_j)$ & $x_s$ & $x_j - x_i  < n_j - m_i$  \\
$(m_i < x_i) \wedge (x_j  \leq n_j)$ & $x_s$ & $x_j - x_i  < n_j - m_i$  \\
$(m_i \leq x_i) \wedge (x_j  \leq n_j)$ & $x_s$ & $x_j - x_i  \leq n_j - m_i$  \\
\hline
$(m_i = x_i) \wedge (x_j  =  n_j)$ & $x_s$ & $x_j - x_i  = n_j - m_i$  \\
\hline
\end{tabular}
\smallskip
\caption{Enabling guard constraints}
\vspace{-18pt}
\label{table:enabling_guard}
\end{table}
%
%
%
%
\subsubsection{Augmenting the Taken Guard.}
For each transition from $q_{s,0}$ to $q_{s+1}$ we augment its guard $g_{s+1}$ by forming $g'_{s+1} = g_{s+1} \wedge tg(\tau_{s,0})$  (see Figure~\ref{fig:bypass1}), where $tg(\tau_{s,0})$ is the {\em taken guard}.
$tg(\tau_{s,0})$ is composed of a single constraint: $0 \leq x_{s,0}$, where $x_{s,0}$ is the clock that is reset at the silent transition $\tau_{s,0}$.
In the next stage of the algorithm of updating the future guards it
will be transformed into the conjunction of the lower bound
constraints $m_i < x_i$ or $m_i \leq x_i$ that appear in $g'_{s,0}$.
These constraints make sure that we spend enough time at $q_{s,0}$ before
moving to the next locations, as if we had taken the silent
transition. The constraint is also used for synchronization of the
future guards in the next step. In Figure \ref{fig:coffee3}, the red-marked part of the
guard from transition $q_{3}$ to $q_6$ shows the taken guard
that has already been updated from $0 \leq x_{2,0}$ to $1<x_1$.
%
%
%
%
\subsubsection{Updating the Future Guards.}
\begin{table}[b]
\centering
\vspace{-5pt}
\begin{tabular}{ l | l | l }
\hline
{\bf Silent Trans. Constr.} & {\bf Future Constr.} & {\bf Replaced Constr.} \\
\hline
$m_i < x_i$, $\{x_{s,0}\}$   \; & $m_{s+j} < x_{s,0}$ \mbox{ or } $m_{s+j} \leq x_{s,0}$  & $m_i + m_{s+j} < x_i$ \; \\
$m_i \leq  x_i$, $\{x_{s,0}\}$   \; & $m_{s+j} < x_{s,0}$  & $m_i + m_{s+j} < x_i$ \; \\
$m_i \leq  x_i$, $\{x_{s,0}\}$   \; & $m_{s+j} \leq x_{s,0}$  & $m_i + m_{s+j} \leq x_i$ \; \\
\hline
$x_i < n_i$, $\{x_{s,0}\}$   \; & $x_{s,0} < n_{s+j}$  \mbox{ or } $x_{s,0} \leq n_{s+j}$  & $x_i < n_i+n_{s+j}$ \; \\
$x_i \leq n_i$, $\{x_{s,0}\}$   \; & $x_{s,0} < n_{s+j}$ & $x_i < n_i+n_{s+j}$ \; \\
$x_i \leq n_i$, $\{x_{s,0}\}$   \; & $x_{s,0} \leq n_{s+j}$  & $x_i \leq n_i+n_{s+j}$ \; \\
\hline
$x_i = n_i$, $\{x_{s,0}\}$ & $x_{s,0} \sim n_{s+j}$  &  $x_i \sim n_i+n_{s+j}$  \\
\hline
\end{tabular}
\smallskip
\caption{Update rules for future guards after removing the silent transitions}
\label{table:updating_guards_1}
\end{table}
The removal of the silent transition $\tau_{s,0}$ enforces updating of
the guards in the paths that start at $q_{s,0}$ and that refer to the
clock $x_{s,0}$, that is reset on the silent transition.
%
The most simple case is when the the silent transition guard $g'_{s,0}$ contains an exact constraint $x_i = n_i$, because then any future constraint of the form $x_{s,0} \sim l$
can be replaced by $x_i \sim n_i+ l$.
%
So, let us assume that the silent transition does not contain an exact constraint.
The rules for updating the future guards are summarized in Table~\ref{table:updating_guards_1}.
Note, that an equality constraint $x_{s,0} = n_{s+j}$ in a future
guard may be treated as $n_{s+j} \leq x_{s,0} \leq n_{s+j}$.
%
%

Let $g_{s+1}, \ldots, g_{s+p}$ be the ordered list of guards of consecutive transitions $\tau_{s+1}, \ldots, \tau_{s+p}$ along a path that starts at $q_{s,0}$.
Then, if $g_{s+j}$ contains the constraint $m_{s+j} < x_{s,0}$, it is replaced by the conjunction of constraints $m_i + m_{s+j} < x_i$, for each constraint $m_i < x_i$ that appear in $g'_{s,0}$.
Similarly, for upper bound constraints. In Figure~\ref{fig:coffee3}, one future guard was updated in the
transition from 
$q_{3}$ to $q_6$: 
The original guard of this
transition was $x_{2,0}=1$ (where $x_{2,0}$ was reset on the silent
transition) and the guard of the silent transition was $1<x_1<2$. Thus,
according to the update rules, the updated future guard is $2<x_1<3$ (written in black), conjuncted
with the taken guard (marked in red).

\begin{wrapfigure}{lb}{50mm}
\vspace{-11pt}
\begin{minipage}{0.18\linewidth}
\begin{tikzpicture}[ta, scale=1.2]
  \node[state] [state,initial] (p1) at (0,0) {$q_0$}; 
  \node[state]  (p2) at (2,0) {$q_1$}; 
  \node[state]  (p3) at (4,0) {$q_2$}; 
  \node[state]  (p4) at (6,0) {$q_3$}; 
  \path [every node/.style={font=\footnotesize}]
      (p1) edge node[above right=1pt and -16pt] [align=left] {$\epsilon$ \\ $1<x_0<2$ \\
        $\{x_{0,1}\}$} (p2)
      (p2) edge node [above] [align=left]  {$\alpha$ \\
        $x_{0,1}=2$ \\ $\{x_1\}$} (p3)
      (p3) edge node [above] [align=left]  {$\alpha$ \\ $x_{0,1}=4$ \\ $\{x_2\}$} (p4);
\end{tikzpicture}
\end{minipage}

\begin{minipage}{0.18\linewidth}
\begin{tikzpicture}[ta, scale=1.2]
  \node[state] [state,initial] (p1) at (0,0) {$q_1$}; 
  \node[state]  (p2) at (2,0) {$q_2$}; 
  \node[state]  (p3) at (4,0) {$q_3$}; 
  \path [every node/.style={font=\footnotesize}]
      (p1) edge node [above right=1pt and -16pt] [align=left]  {$\alpha$ \\
        $3< x_0<4$ \\ $\{x_1\}$} (p2)
      (p2) edge node [above right=1pt and -16pt] [align=left]  {$\alpha$ \\ $5<x_0<6
        \wedge x_1=2$ \\ $\{x_2\}$} (p3);
\end{tikzpicture}
\end{minipage}
\vspace{4pt}
\caption{Guard synchronization}
\label{fig:adexample}
\vspace{-2pt}
\end{wrapfigure}
These rules ensure that each future constraint on the clock $x_{s,0}$ separately conforms to and does not deviate from the possible time range of the silent transition.
Yet, we need to satisfy a second condition: that along each path that
starts at $q_{s,0}$ these future occurrences of $x_{s,0}$ are
synchronized.
This is achieved by augmenting the future guards with
constraints of the form that appear in
Table~\ref{table:updating_guards_2}. No transition in our running
example needs synchronization,
hence we use a different example: the upper
automaton in Figure
\ref{fig:adexample} shows one silent transition followed by two
observable transitions. Using only the previous update rules when
removing the silent transition, the first observable transition might occur between three and four seconds, and the
second one between five and six seconds. If the first transition
occurs after three seconds and the second one after six, this would
not conform to the original automaton which required exactly two seconds between
them. Thus, applying the last synchronization rule of Table~\ref{table:updating_guards_2}, the
constraint $x_1=4-2$ is conjuncted
to the second guard. The lower automaton in Figure \ref{fig:adexample}
illustrates the synchronization. Note, we do not need a bypass
transition here, since the silent transition starts in
the initial state.
\begin{table}[t]
\centering
\begin{tabular}{ l | l | l }
\hline
{\bf Constr. of $g_{s+j}$} & {\bf Constr. of $g_{s+i}$, $\{x_{s+i}\}$, $i<j$} & {\bf Sync. Constr. of $g_{s+j}$} \\
\hline
$m_{s+j} < x_{s,0}$   \; &$x_{s,0} < n_{s+i}$ \mbox{ or } $x_{s,0} \leq n_{s+i}$   & $m_{s+j} - n_{s+i} < x_{s+i}$ \; \\
$m_{s+j} \leq x_{s,0}$   \; &$x_{s,0} < n_{s+i}$ & $m_{s+j} - n_{s+i} < x_{s+i}$ \; \\
$m_{s+j} \leq x_{s,0}$   \; &$x_{s,0} \leq n_{s+i}$ & $m_{s+j} - n_{s+i} \leq x_{s+i}$ \; \\
\hline
$x_{s,0} < n_{s+j}$  \; &  $m_{s+i} < x_{s,0}$  \mbox{ or }  $k_i \leq x_{s,0}$  &  $x_{s+i} < n_{s+j} - m_{s+i}$ \; \\
$x_{s,0} \leq n_{s+j}$  \; &  $m_{s+i} < x_{s,0}$  &  $x_{s+i} < n_{s+j} - m_{s+i}$ \; \\
$x_{s,0} \leq n_{s+j}$  \; &  $m_{s+i} \leq x_{s,0}$  &  $x_{s+i} \leq n_{s+j} - m_{s+i}$ \; \\
\hline
$x_{s,0} = n_{s+j}$  \; &  $x_{s,0} = n_{s+i}$  &  $x_{s+i} = n_{s+j} - n_{s+i}$ \; \\
\hline
\end{tabular}
\smallskip
\caption{Synchronization constraints for future guards after removing silent transitions}
\label{table:updating_guards_2}
\vspace{-18pt}
\end{table}
\subsubsection{Removing the Silent Transition.}
Finally, we can safely remove the silent transition $\tau_{s,0}$ from $q_s$ to $q_{s,0}$ after forming the bypass from $q_{s-1}$ to $q_{s,0}$ with the necessary modifications to the transition guards. 
%
%
\begin{theorem} [Silent Transitions Removal]
\label{th:rst}
 $\LLL(O(A)) = \LLL(A)$.
\end{theorem}
\section{Determinization}
\label{sec:bdtawst}

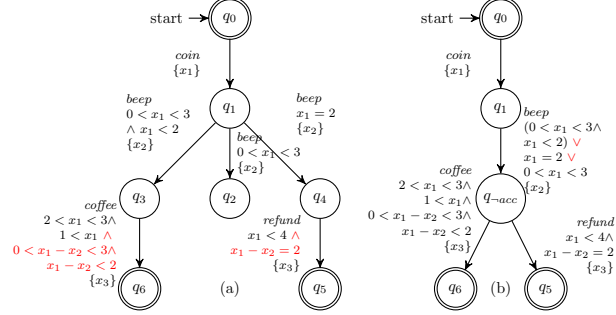
\begin{wrapfigure}{LT}{80mm}
\vspace{-14pt}
\begin{minipage}{0.59\linewidth}
\begin{tikzpicture}[ta, scale=1.2]
  \node[state] [state,initial,accepting] (p1) at (0,0) {$q_0$}; 
  \node[state]  (p2) at (0,-2) {$q_1$}; 
  \node[state]  (p3) at (2,-4) {$q_4$}; 
  \node[state]  (p4) at (-2,-4) {$q_{3}$}; 
  \node[state]  (p5) at (0,-4) {$q_2$}; 
  \node[state,accepting]  (p6) at (2,-6) {$q_5$}; 
  \node[state,accepting]  (p7) at (-2,-6) {$q_6$}; 
  \node at (0,-6) {(a)};
  \path [every node/.style={font=\footnotesize}]
      (p1) edge node[left=15pt] [align=right] {$coin$ \\ 
        $\{x_1\}$} (p2)
  (p2) edge node [above right=8pt and 10pt] [align=left]  {$beep$ \\
        $x_1=2$ \\ $\{x_2\}$} (p3)
      (p2) edge node [above left =-2pt and -7pt] [align=left]  {$beep$ \\ $0 < x_1 < 3$\\
        $\wedge$ $x_1 < 2$ \\ $\{x_2\}$} (p4) 
      (p2) edge node[right=1pt][align=left]  {$beep$ \\ $0 < x_1 < 3$ \\$\{x_{2}\}$} (p5)
      (p3) edge node[left=9pt][align=right]  {{\it refund} \\ $x_1 < 4$  \textcolor{red} {$\wedge
        $} \\  \textcolor{red} {$ x_1-x_2=2$} \\$\{x_3\}$} (p6) 
      (p4) edge node[left = 11pt][align=right]  {{\it coffee} \\ $2<x_1<3 \wedge
        $ \\  $1 < x_1$  \textcolor{red} {$ \wedge$ }\\  \textcolor{red} {$0 < x_1-x_2 < 3
        \wedge$} \\  \textcolor{red} {$x_1-x_2 < 2$}\\ $\{x_3\}$} (p7) ;
\end{tikzpicture}
\end{minipage}%
\begin{minipage}{0.38\linewidth}
\begin{tikzpicture}[ta, scale=1.2]
  \node[state] [state,initial,accepting] (p1) at (0,0) {$q_0$}; 
  \node[state]  (p2) at (0,-2) {$q_1$}; 
  \node[state]  (p4) at (0,-4) {$q_{\neg acc}$}; 
  \node[state,accepting]  (p6) at (1,-6) {$q_5$}; 
  \node[state,accepting]  (p7) at (-1,-6) {$q_6$}; 
  \node at (0,-6) {(b)};
  \path [every node/.style={font=\footnotesize}]
      (p1) edge node[left=15pt] [align=right] {$coin$ \\ 
        $\{x_1\}$} (p2)
      (p2) edge node[right = 11pt][align=left]  {$beep$ \\ $(0 < x_1 < 3
        \wedge$ \\ $x_1 < 2)$ \textcolor{red} {$ \vee $}\\$x_1=2$ \textcolor{red} {$ \vee $} \\$0 < x_1 < 3$  \\  $\{x_2\}$} (p4) 
      (p4) edge node[right=9pt][align=right]  {{\it refund} \\ $x_1 < 4 \wedge
        $ \\ $ x_1-x_2=2$ \\$\{x_3\}$} (p6) 
      (p4) edge node[above left = -8pt and 0pt][align=right]  {{\it coffee} \\ $2<x_1<3 \wedge
        $ \\  $1 < x_1 \wedge$ \\ $0 < x_1-x_2 < 3
        \wedge$ \\ $x_1-x_2 < 2$\\ $\{x_3\}$} (p7) ;
\end{tikzpicture}
\end{minipage}
\vspace{4pt}
\caption{(a) Modified guards added to future transitions (b) determinization via
  disjunction} \label{fig:coffee4}
\vspace{2pt}
\end{wrapfigure}

Existing determinization algorithms (as e.g. applied in \cite{tacas14}) create the powerset of
all transitions to be determinized, and build one transition
for each subset in the powerset. We propose an alternative approach, that
reduces the amount of locations and transitions in the deterministic
automata, by shifting some complexity towards the guards. 
Our motivation is the use of SMT solvers for verifying the timed
automata models.
The larger guards can be directly converted into SMT-LIB formulas, and
thus should not pose a problem.

The approach works under the following prerequisites: After the removal
of the silent transitions the timed automaton $A$ is
in the form of a tree of depth $k$. At each level $i$ the same new clock $x_i$ is reset on each of the
transitions of that level. This is the only clock reset
on this level, and no clock is ever reset again.

The basic idea behind the determinization algorithm is to merge all
transitions of the same source location and the same action via disjunction, and to push the decision which of them was actually taken to the following transitions.
The postponed decision which transition was actually taken can be
solved later on by forming diagonal constraints (as in zones) that are
invariants of the time progress, and are conjuncted 
to immediately following transitions. Note that the distinction between
  accepting and non-accepting locations increases complexity slightly:
  the determinization of transitions leading to accepting locations
  and transitions leading to non-accepting locations can not be done
  exclusively by disjunction of their guards. We therefore need to add an accepting and a non-accepting
  location to the deterministic tree, and merge all transitions
  leading to non-accepting locations and
  all transitions leading to accepting locations separately. To ensure
  determinism for these
  transitions, we conjunct the negated guard
  of the accepting transition to the guard of the non-accepting
  transition. 

A pseudo-code description is given in Algorithm~\ref{alg:godet}. The
determinization is done in several steps applied to every
location $q$ with multiple outgoing transitions with the same action  (Line \ref{line:det_exi}),
starting at the initial location (Line~\ref{line:det_init1}). 
Let $q_i$ be such a location with multiple
$\alpha$ transitions (Line \ref{line:det_eachtau}). First, we add an
accepting  and a non-accepting location $q_ {acc}$, $q_{\neg acc}$ replacing the target locations of the multiple $\alpha$ transitions
(Line \ref{line:det_newloc}). Then, for each $\tau_i$ in the $\alpha$
transitions with guard $g$ from $q_i$ to $q_{i+1}$, let $g'$ be the
result of subtracting the clock $x_{i+1}$ that is reset on $\tau_i$ from
all clocks that appear in $g$ (Lines \ref{line:det_substart}-\ref{line:det_subend}).
Next, $g'$ is conjuncted to the guards of each transition $\tau_{i+1}$ that
follows $\tau_i$ and the source location of $\tau_{i+1}$ is set to
either $q_{acc}$ or $q_{\neg acc}$, depending on whether $q_{i+1}$ is accepting or not. Transitions leaving $q_{\neg acc}$
are additionally copied to $q_{acc}$, in case the guards of $\alpha$
transitions overlap. (Lines
\ref{line:det_addaccgo},\ref{line:det_addnaccgo}). Note that $g'$ evaluates to $true$ in every
branch below $\tau_i$ if $\tau_i$ was enabled, thus the conjunction does not change the language
of the automaton. Figure~\ref{fig:coffee4}(a) illustrates the conjunction of the
modified guards on our running example, marked in red. Note that the determinization
did not involve any accepting locations, thus there was no splitting
into $q_{acc}$ and $q_{\neg acc}$. 
Next, all the $\alpha$-transitions from $q$ leading to
accepting locations are merged into
a transition leading to $q_{acc}$ (Line \ref{line:det_addtransacc})
and all others into a transition leading to $q_{\neg acc}$(Line \ref{line:det_addtransnacc}), by disjuncting their guards
(Lines  \ref{line:det_gacc},\ref{line:det_gnacc}).
The guard of the transition leading to $q_{\neg acc}$ is conjuncted to
the negation of the other guard, to ensure determinism (Line
\ref{line:det_addtransnacc}). Finally, all merged $\tau_i$ and their
target locations can be removed (Line \ref{line:det_remove1}).
Figure~\ref{fig:coffee4}(b) shows the determinized
coffee-machine. 

\algnewcommand{\IIf}[1]{\State\algorithmicif\ #1\ \algorithmicthen}
\algnewcommand{\EndIIf}{\unskip\ \algorithmicend\ \algorithmicif}
\begin{algorithm}[H]
\begin{algorithmic}[1]
\Require $A \in\nta_k$ in the form of a tree of depth $k$ with renamed clocks
\Ensure $D(A) \in \ta_k$, such that $\LLL(D(A)) = \LLL(A)$

\State $P \leftarrow \{(\Qinit,0)\}$ \label{line:det_init1}
\While{$P \neq \emptyset$} \label{line:det_while}
	\State \textsc{Pick} $(q_i, i) \in P$; $P \leftarrow P \backslash (q_i, i)$
      \For{each $\alpha \in \Actions$}
           \If {$\exists\ \tau_1 ( q_i,\alpha, g_1, \{x_{i+1} \}, q_{1}) \neq \tau_2
                                         ( q_i ,\alpha, g_2, \{x_{i+1} \}, q_{2})$} \label{line:det_exi}
                \State $g_{acc} \leftarrow false$; $g_{\neg acc} \leftarrow false$
                \State \textsc{Add} new locations $q_{acc}$, $q_{\neg acc}$ \label{line:det_newloc}
                \For{each transition $\tau_{i} ( q_i, \alpha,
                  g_{i+1}, \{x_{i+1} \}, q_{i+1})$}  \label{line:det_eachtau}
                        \State $g' \leftarrow g_{i+1}$  \label{line:det_substart}
                        \For{each clock $x_j$ in $g_{i+1} $}
                                  \State $ g' \leftarrow g' [x_j := x_j - x_{i+1}]$ \label{line:det_diag}
                        \EndFor \label{line:det_subend}
                        \For{each transition \label{line:det_addgostart}
                                           $\tau_{i+1} (q_{i+1},\beta,g_{i+2},\{x_{i+2}\},q_{i+2})$}
                               \State \textsc{Add} $\tau_{acc} (q_{acc},\beta,
                                           (g_{i+2} \wedge g')
                                           , \{x_{i+2}\}, q_{i+2})$ \label{line:det_addaccgo}
                               \State \textsc{Add} $\tau_{\neg acc} (q_{\neg acc},\beta,
                                           (g_{i+2} \wedge g')
                                           ,\{x_{i+2}\}, q_{i+2})$ \label{line:det_addnaccgo}
                                            \State \textsc{Remove} $\tau_{i+1}$
                               \EndFor \label{line:det_addgoend}
                                         
                                         \IIf{$accepting(q_{i+1})$} $g_{acc} \leftarrow g_{acc} \vee g_{i+1}$  \EndIIf \label{line:det_gacc}  
                                         \IIf{$\neg
                                           accepting(q_{i+1})$}
                                         $g_{\neg acc} \leftarrow
                                         g_{\neg acc} \vee g_{i+1}$  \EndIIf  \label{line:det_gnacc}
 
                                         \State \textsc{Remove} $\tau_{i}$
                                         and $q_{i+1}$ \label{line:det_remove1}
                               \EndFor
                                      \State \textsc{Add} transition $\tau_{acc} (q_i,\alpha, g_{acc}, \{x_{i+1} \}, q_{acc})$ \label{line:det_addtransacc}
                                           \State \textsc{Add} transition $\tau_{\neg acc} (
                               q_i,\alpha, (g_{\neg acc} \wedge \neg g_{acc}), \{x_{i+1}\},
                               q_{\neg acc})$ \label{line:det_addtransnacc}
                 \EndIf
         \EndFor
         \For{each transition $\tau_{i} ( q_i,\alpha, g_{i+1}, \{x_{i+1}\}, q_{i+1})$}
                 \State $P \leftarrow P \cup (q_{i+1},i+1)$ \label{line:plusp}
         \EndFor
\EndWhile
\end{algorithmic}
\caption{Guard-Oriented Determinization}
\label{alg:godet}
\end{algorithm}

\begin{theorem}[Determinization]
\label{th:GOdet}
The determinization algorithm constructs a deterministic timed automaton $D(A)$ such that $\LLL(D(A)) = \LLL(A)$.
\end{theorem}

\section{Complexity}
\label{sec:complexity}
{\bf Bounded Unfolding.}
We unfold the timed automaton $A$ into a tree and cut it when reaching observable level $k$.
Let us assume that the tree is of depth $K$, $K \geq k$, and of size $N = O(d^{K})$, with $d \geq 1$ representing the approximate out-degree of the vertices in the graph of $A$.
Since the analysis of the SMT solvers for different applications requires the exploration of all the transitions in the unfolded graph of $A$, the unfolding stage of our algorithm does not necessarily increase the overall time complexity of the algorithm.
%
%
\\
{\bf Removing Silent Transitions.} 
Our algorithm does not increase the size of the tree since we only substitute the silent transitions by the bypass transitions.
We do add, however, constraints.
The number of enabling-guard constraints that we add to each bypass transition is of order $O(K^2)$.
Each updated future constraint is of order $O(K)$ (including on-the-fly simplification, so that each clock has at most one lower and one upper bound), and each future transition may be updated at most $O(K)$ times. Hence, the updating step is also of order $O(K^{2})$, and the complexity of the whole algorithm is $O(N K^2)$.
%
%
%
%
%
Note, we do not need to transform the diagonal constraints introduced in the algorithm into unary constraints, nor do they introduce problems in the next algorithm of determinization.
\\
{\bf Determinization}
%
decreases the size of the unfolded automaton, if non-\-de\-ter\-min\-ism exists. 
The complexity gain can be exponential in the number of locations and transitions, but is lost by a proportional larger complexity in the guards.
%
%

%
%
%
%
\section{Implementation and Experimental Results}
\label{sec:exp_res}
The algorithms were implemented in Scala (Version 2.10.3) and integrated into the 
test-case generation tool MoMuT::TA\footnote{\url{https://momut.org/?page_id=355}}, providing a significant increase in the capabilities of the tool. 
MoMuT::TA provides model-based mutation testing algorithms for timed automata~\cite{DBLP:conf/tap/AichernigLN13}, using UPPAAL's~\cite{UPPAAL} XML format as input and output.
%
The determinization algorithm use the SMT-solver Z3~\cite{z3} for
checking satisfiability of guards. 
All experiments were run
on a MacBook Pro with a 2.53 GHz Intel Core 2
Duo Processor and 4 GB RAM.

The implementation is still a prototype and further 
optimizations are planned.
One already implemented optimization is the "on-the-fly"
execution of the presented algorithms, allowing the unrolling,
clock renaming, silent transition removal and determinization in one
single walk through the tree. The combined algorithm does not suffer
from the full exponential blow-up of the unfolding: if the automaton
contains a location that can be reached via different traces, yet with
the same clock resets, the 
unfolding splits it into several, separately processed, locations, while the on-the-fly algorithm only needs to process it once.

The following studies 
compare the
numbers of locations and the runtimes of $a)$ the silent transition removal, $b)$ a standard
determinization algorithm that works by splitting non-deterministic
transitions into several transitions that contain each possible
combination of their guards, $c)$ the new determinization algorithm
introduced in Section \ref{sec:bdtawst} and $d)$ its  on-the-fly
version.
%
\begin{figure*}[t]
\begin{center}
\begin{minipage}{0.22\linewidth}
\begin{tikzpicture}[ta, scale=1.2]
  \node[state] [state,initial,accepting] (p1) at (-1,-1) {};
  \node[state]  (p2) at (1,-1){} ;
  \node at (0,-5.3) {(a)};
  \path [every node/.style={font=\footnotesize}]
      (p1) edge[bend  left] node[above] [align=left] {$\beta$ \\$0 < x  < 1$} (p2)
      (p1) edge[loop below] node [below=10pt][align=left]  {$\alpha$ \\
        $x=1$ \\ $\{x\}$} (p1)
      (p2) edge[bend  left] node[below][align=left]  {$\epsilon$ \\ $x=1$ \\ $\{x\}$ } (p1);
\end{tikzpicture}
\end{minipage}
\begin{minipage}{0.22\linewidth}
\begin{tikzpicture}[ta, scale=1.2]
  \node[state] [state,initial,accepting] (p1) at (-1,-1) {};
  \node[state]  (p2) at (1,-1){} ;
  \node at (0,-5.3) {(b)};
  \path [every node/.style={font=\footnotesize}]
      (p1) edge[bend  left] node[above] [align=left] {$\beta$ \\$0 < x < 1$} (p2)
      (p1) edge[loop below] node [below=10pt][align=left]  {$\alpha$ \\
        $x=1$ \\ $\{x\}$} (p1)
      (p2) edge[bend  left] node[below][align=left]  {$\epsilon$ \\
        $x=1$ \\ $\{x\}$ } (p1)
      (p2) edge[loop below] node [below=10pt][align=left]  {$\alpha$ \\
        $x=1$ \\ $\{x\}$} (p2);
\end{tikzpicture}
\end{minipage}
\begin{minipage}{0.22\linewidth}
\begin{tikzpicture}[ta, scale=1.2]
  \node[state] (p1) at (-1.5,0) {};
  \node[state,initial,initial where=above]  (p2) at (1.5,0) {};
  \node[state]  (p3) at (-1.5,-3) {};
  \node[state,accepting]  (p4) at (1.5,-3) {};
  \node at (0,-4) {(c)};
  \path [every node/.style={font=\footnotesize}]
      (p1) edge[bend left] node[above] [align=right] {$\alpha$ \\ $x>0$ \\ $\{x\}$} (p2)
      (p2) edge node[below left] [align=left] {$\alpha$ \\ $x>0$ \\ $\{x\}$} (p1)
      (p2) edge node[below right] [align=left] {$\alpha$ \\ $x>0$  } (p3)
      (p3) edge node[below] [align=left] {$\alpha$ \\ $x=1$} (p4) ;
\end{tikzpicture}
\end{minipage}
\begin{minipage}{0.22\linewidth}
\begin{tikzpicture}[ta, scale=1.2]
  \node[state] (p1) at (-1.5,0) {};
  \node[state,initial,initial where=above]  (p2) at (1.5,0) {};
  \node[state]  (p3) at (-1.5,-3) {};
  \node[state,accepting]  (p4) at (1.5,-3) {};
  \node at (0,-4) {(d)};
  \path [every node/.style={font=\footnotesize}]
      (p1) edge[bend left] node[above] [align=right] {$\alpha$ \\ $x>0$ \\ $\{x\}$} (p2)
      (p2) edge node[below left] [align=left] {$\alpha$ \\ $x>0$ \\ $\{x\}$} (p1)
      (p2) edge node[below right] [align=left] {$\alpha$ \\ $x>0$  }
      (p3)
      (p4) edge node[right] [align=left] {$\epsilon$ \\ $1 < x <
        3$ \\ $\{x\}$  } (p2)
      (p3) edge node[below] [align=left] {$\alpha$ \\ $x=1$} (p4) ;
\end{tikzpicture}
\end{minipage}
\end{center}
\caption{The four timed automata used in Study 1 and Study 2} 
\label{fig:results}
\end{figure*}
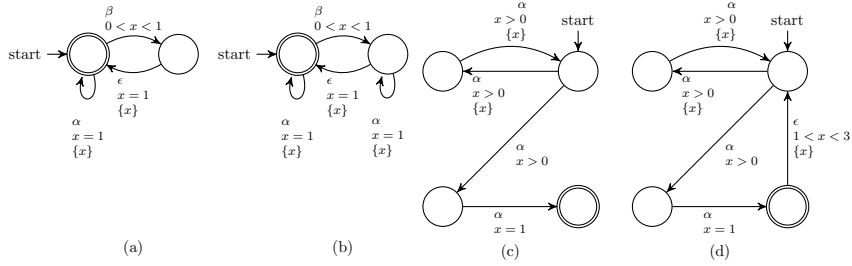
\\
{\bf Study 1.}
The first example, taken from Diekert et al.~\cite{DiekertGP97}, is
the timed automaton illustrated in Fig.~\ref{fig:results} (a), which
cannot be determinized.
We then added another $\alpha$-transition (Fig.~\ref{fig:results} (b)), which causes non-determinism after removing the silent transition.
The test results are shown in Table \ref{tbl:a1_results} (before and
after modification).
%
\begin{table}[b]
\centering
\setlength{\tabcolsep}{0.2mm}			
  \begin{tabular}{ c | r | r | r | r | r | r |r | r}
Depth & \multicolumn{4}{c | }{Number of locations} &
\multicolumn{4}{c}{Runtime (sec.)}\\
\hline
 & unfolded & std.~det. & new det. & on-the-fly
 & $\epsilon$-removal & std.~det. & new det. & on-the-fly \\
\hline
2 & 8 & 7 & 7 & 7 & 0.1 & 0.3 & 0.1 & 0.1 \\
5 & 78 & 63 & 63 & 63 & 0.4 & 0.5 & 0.4 & 0.2\\ 
9 & 1,278 &  1,023 & 1,023 & 1,023 & 16,011.2 & 6.7 & 7.2 &  1.0 \\ 
\hline
2 & 9 & 8 & 8 & 8 & 0.2 & 0.2 & 0.2 & 0.1 \\
5 & 177 & 135 & 84 & 63 & 0.8 & 0.9 & 1.3 & 0.7 \\ 
9 & 8,361 &  4,364 & 3,609 & 1,023 & 20,969.0 & 71.2 & 88.3 & 9.6 \\ 
    \end{tabular}
\caption{Runtime and number of locations for the automata of
  Fig.~\ref{fig:results} (a) (first three rows) and
  Fig.~\ref{fig:results} (b) (last three rows)}
\label{tbl:a1_results}
\end{table}
%
\\
{\bf Study 2.}
The second example is taken from Baier et al.~\cite{bbbb} and is illustrated in Fig.~\ref{fig:results}(c).
We modified the automaton by adding a silent transition (Fig.~\ref{fig:results}(d)).
Table~\ref{tbl:a3_results} shows the results of the two determinization approaches.
%
%
%
\begin{table}[t]
\centering
\setlength{\tabcolsep}{0.2mm}			
  \begin{tabular}{ c | r | r | r | r | r | r |r | r}
Depth & \multicolumn{4}{c | }{Number of locations} &
\multicolumn{4}{c}{Runtime (sec.)}\\
\hline
 & unfolded & std.~det. & new det. & on-the-fly
 & $\epsilon$-removal & std.~det. & new det. & on-the-fly \\
\hline
2 & 5 & 5 & 4 & 4 & - & 0.1 & 0.1 & 0.1\\
5 & 11 & 10 & 8 & 8 & -  & 0.2 & 0.3 & 0.1 \\ 
10 & 21 &  21 & 16 & 16 & - & 0.3 & 0.3 & 0.1\\
 25 & 51 &  50 & 38 & 38 & - & 0.5 & 0.9 & 0.2 \\ 
 50 & 101 &  100 & 76 & 76 & - & 0.7 & 391.6 &  0.3 \\
\hline
2 & 5 & 5 & 4 & 4 & 0.1 & 0.1 & 0.1 & 0.01\\
5 & 24 & 26 & 8 & 8 & 0.2 & 2.1 & 0.4 & 0.3\\ 
10 & 140 &  661 & 16 & 16 & 0.5 & 1,945.1 & 2.1 & 0.5 \\ 
\end{tabular}
\caption{Runtime and number of locations for the automata of Fig.~\ref{fig:results} (c) (first three rows) and Fig.~\ref{fig:results} (d) (last three rows)}
\label{tbl:a3_results}
\vspace{-5pt}
\end{table}
\\
{\bf Study 3.}
This study is part of a model of an industrial application: it is based on a car alarm system that was already used as an example in our work on model-based mutation testing from timed automata~(see \cite{DBLP:conf/tap/AichernigLN13} for the whole model).
In this evaluation, we introduced a silent transition that adds a non-deterministic delay of up to two seconds before the timer of the alarm starts,
and our results are given in Table \ref{tbl:cas_results}.
We were able to perform the removal of silent transitions and the
guard-oriented determinization up to depth 12, and the
location-oriented determinization up to depth 8.

As expected, the studies confirm that the complexity of the different algorithms
depends vastly on the input models. For the current paper we picked
two small examples that were introduced in previous papers on determinization
and one example that was an industrial use case in a previous
project. Our next step will be a stronger evaluation on a larger case
study. 
The tool and the current examples are available\footnote{\url{https://momut.org/?page_id=394}}.
\begin{table}[b]
\vspace{-2pt}
\centering
\setlength{\tabcolsep}{0.2mm}			
  \begin{tabular}{ c | r | r | r | r | r | r |r | r}
Depth & \multicolumn{4}{c | }{Number of locations} &
\multicolumn{4}{c}{Runtime (sec.)}\\
\hline
 & unfolded & std.~det. & new det. & on-the-fly
 & $\epsilon$-removal & std.~det. & new det. & on-the-fly \\
\hline
2 & 8 & 8 & 8 & 8 & 0.108 & 0.2 & 0.1 & 0.0\\
5 & 153 & 139 & 83 & 81 & 0.4 & 1.0 & 0.8 & 0.2 \\ 
8 & 2,062 & 1,973 & 757 & 739 & 4.1 & 129.0 & 11.6 & 0.9 \\
12 & 78,847 &  - & 14,009 & 13,545 & 10,592.3 & - & 4,832.1 & 10.2\\
\end{tabular}
\caption{Runtime and number of locations for the Car Alarm
  System~\cite{DBLP:conf/tap/AichernigLN13}, modified by adding a
  silent transition causing a 0-2 seconds delay.}
\label{tbl:cas_results}
\end{table}
\section{Related Work}
\label{sec:related_work}

The main inspiration to our work comes from~\cite{ta-eps} and~\cite{bbbb}. B\'{e}rard et al.~\cite{ta-eps} show that 
silent transitions extend the expressive power of TA and identify a sub-class of eNTA for which silent transitions can be removed. 
By restricting our selves to the bounded setting, we can remove silent transition of all strongly-responsive eNTAs. In addition, 
our approach for removing silent transitions preserves diagonal constraints in the resulting automaton, thus avoiding a potential 
exponential blow-up in the size of its representation (see~\cite{bouyer2} for the practical advantages of preserving diagonal 
constraints in TA). Baier et al.~\cite{bbbb} propose a procedure for translating $\nta$ to {\em infinite} DTA trees, 
and then identify several classes of $\nta$ that can be effectively determinized into finite DTA.
In contrast to our work, their procedure works on the region graph, which makes it impractical for implementation.
In addition, we also allow in our determinization procedure disjunctive constraints which results in a more succint representation that 
can be directly handled by the bounded model checking tools.
Both~\cite{ta-eps} and \cite{bbbb} tackle non-determinism and observabilty in TA from a general theoretical perspective. 
We adapt the ideas from these papers 
and propose an effective procedure for the bounded determinization of $\ntaeps$.

Wang et. al~\cite{tacas14} use timed automata for language inclusion. Their procedure involves building a tree, renaming the clocks and determinization of the tree. Contrary to our work, they do not restrict themselves to the bounded setting, thus taking the risk that their algorithm does not terminate for some classes of timed automata. Also, they use the "standard" determinization method that involves splitting non-deterministic transitions into a possibly far larger set of deterministic transitions, whereas we join them into one transition.

Krichen and Tripakis~\cite{tripakis-testing} produce deterministic testers for non-deterministic timed automata in the context of model-based testing.  
They restrain the testers to using only one clock, which is reset upon receiving an input. The testers are sound, but not in general complete and might accept behavior of the system under test that should be rejected. Bertrand et al.~\cite{bertrand} develop a game-based method for determinization of eNTA which generates either a language equivalent DTA when 
possible, or its approximation otherwise. A similar approach is proposed in~\cite{bertrand2011} in the context of model-based testing, where it is 
shown that their approximate determinization procedure preserves the tioco relation. In contrast to our approach, which is language preserving up to a 
bound $k$, and thus appropriate for bounded model checking algorithms, determinization in the above-mentioned papers introduces a different kind of 
approximation than ours.


\section{Conclusion}
\label{sec:conclusion}
The bounded setting allows the handling of a larger class of TA and in a more efficient way than in the unbounded setting.
The extension from standard unary constraints to diagonal and disjuncive constraints has a practical reason: it is more efficient to let the SMT solvers deal with them than to translate them into standard form.
In this paper a novel procedure was presented, which transforms bounded, non-deterministic and partially-observable TA into deterministic and fully-observable TA with diagonal and disjunctive constraints.
The procedure includes an algorithm for removing the silent transitions and a determinization algorithm.
It was implemented, tested and integrated into a model-based test generation tool. Recently~\cite{amost15} we investigated ways of pruning the determinized tree, to reduce the state space of the unfolding. These appoaches look promising for applying the presented work to test-case generation in industrial studies.
\vspace{-4pt}
%
%

\paragraph{Acknowledgement.}
The research leading to these results has received funding from the
ARTEMIS Joint Undertaking under grant agreement N\textordmasculine\,332830 and from the Austrian Research Promotion
Agency (FFG) under grant agreement N\textordmasculine\,838498 for the
implementation of the project CRYSTAL, Critical System Engineering Acceleration. 

\bibliographystyle{plain}
\bibliography{ref}
%
%
\section{Appendix A - Renaming of Clocks}
\label{sec:rename_clks}
\begin{figure}[H]
\vspace{-15pt}
\begin{minipage}[t]{1\linewidth}
\begin{algorithm}[H]
\begin{algorithmic}[1]
\Require $A \in \ntaeps_K$, a tree of depth $K$ and observable depth $k$, clocks $\Clocks$, $| \Clocks | = n$
\Ensure $A \in \ntaeps_K$, clocks $\Clocks'$,  $| \Clocks' | = K$, single clock reset per transition, same clock reset at same (observable, silent) level
\State $l_1 \leftarrow 0$
\Comment observable (primary) level
\State $l_2 \leftarrow -1$
\Comment silent (secondary) level
\For{$i \leftarrow 0,..,n-1$}
	\State $X[i] \leftarrow x_0$
	\Comment $x_0$ is reset at the initial location
\EndFor
\State \Call{RenameClocks}{$q_0, X, l_1, l_2$}
\Procedure{RenameClocks}{$q, X, l_1, l_2$}
\For{each $\tau = (q, \alpha, g, \ResetClocks, q') \in trans(q)$} 
	\For{$i \leftarrow 0,..,n-1$}
		\State $g \leftarrow g[x_i \leftarrow X[i]]$
		\Comment renaming the clocks in the guard $g$
	\EndFor
	\If{$\alpha = \epsilon$}
	\Comment silent transition
		\State $l_2 \leftarrow l_2 + 1$
		\State $x \leftarrow x_{l_1, l_2}$
		\Comment the new reset clock in case of a silent trans.
	\Else   
		\State $l_1 \leftarrow l_1 + 1$
		\State $l_2 \leftarrow -1$
		\State $x \leftarrow x_{l_1}$
		\Comment  the new reset clock in case of an observable trans.
	\EndIf
	\For{$i \leftarrow 0,..,n-1$}
		\If{$x_i \in \ResetClocks$}
			\State $X[i] \leftarrow x$
			\Comment updating the clock substitution list
		\EndIf
	\EndFor
	\State $\ResetClocks \leftarrow \{ x \}$
	\Comment updating the reset clocks of $\tau$
	\If{$l_1 < k$}
		\State \Call{RenameClocks}{$q', X, l_1, l_2$}                
		\Comment recursive call with the target location
	\EndIf
\EndFor
\EndProcedure
\end{algorithmic}
\caption{Renaming the Clocks}
\label{alg:rename_clks}
\end{algorithm}
\end{minipage}
\end{figure}
%
The concrete algorithm used for renaming of the clocks is presented in pseudo-code in Algorithm~\ref{alg:rename_clks}.
The original clocks are $x_0, \ldots, x_{n-1}$.
Each new clock has either one index ($l_1$) in case the transition in which it is reset is observable, or two indices ($l_1, l_2$) in case of a silent transition.
After the removal of the silent transitions stage we will be left with clocks with a single index and the same clock reset for the same level of the tree.
The vector $X[0..n-1]$ holds the clock substitution list: $X[i]$ refers to the new clock that substitutes the original clock $x_i$.
The the set of transition with source location $q$ is denoted by $trans(q)$.
\section{Appendix B - Proofs}
\label{sec:proofs}
%
%
%
%
%
%
\subsection{Proof of Theorem~\ref{th:rst} [Silent Transitions Removal]}
\label{proof_th:rst}
Given a non-deterministic timed automaton with silent transitions $A$ in the form of a finite tree, we need to show that our algorithm of removing the silent transitions results in an equivalent timed automaton, that is, $\LLL(O(A)) = \LLL(A)$.
That is, we will show that if $A'$ is the result of removing one first silent transition then $A$ and $A'$ are equivalent: for every timed trace of $A$ there is an equivalent timed trace of $A'$ and vice versa, in the sense that the corresponding observable timed traces are identical.  

We claim that by induction the proof of equivalence for a single removal of a first silent transition suffices to prove the theorem.
First, there are only finitely-many silent transitions in $A$.
Secondly, the removal of a silent transition does not change the form of the guards at the part of the automaton that contains the remaining silent transitions: the introduction of diagonal constraints happens only at the enabling guard and so the algorithm for removal of the next silent transitions remains the same.

So, let $\tau_{s,0}$ be a first silent transition on a path $\gamma$ that starts at the initial location.
Let $\tau_{s,0}$ be from location $q_s$ to location $q_{s,0}$, let $q_{s-1}$ be the location that leads to $q_s$ and let $q_{s+1}$ be a location that follows $q_{s,0}$ on the path.
Let $A'$ be the automaton that results after removing $\tau$ and performing the steps as in Algorithm~\ref{alg:rst}.
Clearly, for every run that does not pass through $\tau_{s,0}$ there is an identical run in the other automaton.
Thus, we restrict ourselves to runs though $\tau_{s,0}$.
\subsubsection{$\LLL(A) \subseteq \LLL(A')$.}
Let $\rho$ be a run on $A$ through $\gamma$.
We need to show that there exists a run $\rho'$ on $A'$ with the same observable trace as of $\rho$.
The run $\rho'$ will go through the same locations and transitions as does $\rho$, except for the part $q_{s-1}$, $\tau_s$, $q_s$, $\tau_{s,0}$, $q_{s,0}$ in $A$ which will be replaced by the bypass $q_{s-1}$, $\tau'_s$, $q_{s,0}$ in $A'$ as in Fig.~\ref{fig:bypass1}.
The dates of the transitions will also be the same, except for the silent transition that is missing in $\rho'$.
That is, if $t_s$, $t_{s,0}$ and $t_{s+1}$ are the dates of $\rho$ at the transitions $\tau_s$, $\tau_{s,0}$ (the silent transition) and $\tau_{s+1}$ then the corresponding transitions of $\rho'$ will take place at $t_s$ (the time of the bypass) and $t_{s+1}$.

Since $\rho$ goes through $\tau_{s,0}$, we know that by the time $t_s$ after the reset of clock $x_s$ the guard $g_{s,0}$ of $\tau_{s,0}$ is satisfied in some non-negative time.
Thus, we know that each constraint of a clock $x_j$ that appears in $g_{s,0}$ is satisfied at a non-negative delay, and that all these constraints can be satisfied simultaneously.
So, first we need to show that the corresponding guard $g'_s = g_s \wedge eg(\tau_{s,0})$ of $\tau'_s$ in $\rho'$ is satisfied at the same time, that is, that the enabled guard $eg(\tau_{s,0})$ is satisfied at $t_s$.

We will mostly restrict ourselves to strict inequalities, as the extension to the other cases (strict inequality versus weak inequality or weak inequality versus weak inequality) is straight forward.

For each clock $x_j$ that is not reset at $\tau_s$ (that is, $j \neq s$) and that appears with an upper bound constraint $x_j < n_j$ (or $x_j \leq n_j$) at $g_{s,0}$ clearly the same constraint holds also at the not-later time $t_s$.
But that part is exactly what we have in $eg(\tau_{s,0})$ when comparing the upper bound constraint of $x_j$ with the lower bound constraint of the reset clock $x_s$.
Here the constraint in $eg(\tau_{s,0})$ is, in general, $x_j - x_s < n_j - m_s$, and  since $m_s = 0$ and $x_s$ is reset at $\tau'_s$ and replaced by $0$ in the inequality the result is indeed $x_j < n_j$.

In addition to the above unary constraints, we know that each upper bound constraint on clock $x_j$ in $g_{s,0}$ refers to a time which is of greater delay than the delay needed to reach each lower bound constraint on clock $x_i$ in $g_{s,0}$, that is, $n_j - x_j > m_i - x_i$ at time $t_{s,0}$, otherwise these constraints couldn't have been satisfied simultaneously in $\rho$.
But this is indeed the constraint $x_j - x_i < n_j - m_i$ that appears in $eg(\tau_{s,0})$.

We have seen that all the constraints of $eg(\tau_{s,0})$ are satisfied at time $t_s$ and so the constraint $g'_s$ of $\rho'$ is satisfied at $t_s$ and the transition $\tau'_s$ can be taken.

The next step is to show that the transition $\tau_{s+1}$ with guard $g'_{s+1}$ of $\rho'$ from location $q_{s,0}$ to location $q_{s+1}$, as well as the next transitions $\tau_{s+j}$, $j = 2, \ldots, p$, with guards $g'_{s+j}$ can be taken at the same dates $t_{s+j}$ on which $\tau_{s+j}$ are taken in $\rho$ on guards $g_{s+j}$, $j = 1, \ldots, p$. 

If the silent transition happens to be on an exact time: $x_i = n_i$ then the update of the future guards that refer to the clock $x_{s,0}$ that was reset at $\tau_{s,0}$ is clear: each occurrence of $x_{s,0}$ is replaced by $x_i - n_i$, and we are done.
So, suppose that there are no exact constraints at the silent transition.

For simplicity we will restrict ourselves mostly to strict inequalities and write the guard $g'_{s,0}$ of the silent transition $\tau_{s,0}$ as:
\begin{equation}
g'_{s,0} = 0 \leq x_s \wedge \bigwedge_{i=2, \ldots, r} m_i < x_i < n_i,
\end{equation}
where for some of the clocks $x_i$ there may be only a lower bound or only an upper bound constraint.

The constraints on $x_{s,0}$ at the transitions $\tau_{s+j}$, $j = 1, \ldots, p$ contain $0 \leq x_{s,0}$ in $\tau_{s+1}$ and are of the general (strict inequalities) form $m_{s+j} < x_{s,0} < n_{s+j}$ in $\tau_{s+j}$.
The corresponding updated constraints of $A'$ at time $t_{s+j}$, $j = 1, \ldots, p$, are
\begin{equation}
\label{eq:updated_guard}
\bigwedge_{i=1, \ldots, r} m_i +m_{s+j} < x_i < n_i + n_{s+j}.
\end{equation}

First, we need to show that the taken guard $tg(\tau_{s,0})$ is satisfied at time $t_{s+1}$.
The taken guard is the constraint $0 \leq x_{s,0}$.
After the update of the future guards this constraint is replaced by the conjunction of all the lower bound constraints $m_i < x_i$ of $g'_{s,0}$.
But since these lower bound constraints are satisfied at the time $t_{s,0}$ of the silent transition (in $\rho$) then clearly they are satisfied at $t_{s+1}$, $t_{s+1} \geq t_{s,0}$, that is, the updated taken guard $tg(\tau_{s,0})$ is satisfied in $\rho'$.

Let us look at the other updated future constraints.
Since at the time of the silent transition $x_{s,0} = 0$ and $m_i < x_i$ then at time $t_{s+j}$ when $m_{s+j} < x_{s,0}$ we have $m_i +m_{s+j} < x_i$.
With a similar argument for the upper bound constraints, we see that the constraints of (\ref{eq:updated_guard}) are satisfied in $\rho'$..

Also the part of the synchronization rules is clear since it refers to the possible minimum and maximum time difference between every two transitions on which $x_{s,0}$ occurs, and since the run $\rho$ goes through these transitions it assures that these constraints can be satisfied.
So, for example, the synchronization constraint $m_{s+j} - n_{s+i} < x_{s+i} < n_{s+j} - m_{s+i}$ that is added to the guard $g_{s+j}$ of $\tau_{s+j}$, refers to the time difference $t_{s+j} - t_{s+i}$ between the transition $\tau_{s+i}$ and the transition $\tau_{s+j}$, $i < j$.

Note that the synchronization with the constraint $0 \leq x_{s,0}$ of $\tau_{s+1}$ results in adding to $\tau_{s+j}$, $j = 1, \ldots, p$ the constraint $x_{s+1} < n_{s+j}$, that is $t_{s+j} - t_{s+1} < n_{s+j}$, which clearly is satisfied since $t_{s+j} - t_{s,0} < n_{s+j}$.

We showed that the observable trace of $\rho'$ is the same as that of $\rho$ and this completes the proof of $\LLL(A) \subseteq \LLL(A')$.
\subsubsection{$\LLL(A') \subseteq \LLL(A)$.}
Let $\rho'$ be a run on $A'$ going through the bypass $\tau'_s$.
We will show that there exists a run $\rho$ through $\tau_{s,0}$ in $A$ with the same observable trace as of $\rho'$.

The first thing we need to check is that the silent transition $\tau_{s,0}$ can be taken, given that the enabling guard $eg(\tau_{s,0})$ is satisfied at time $t_s$.
The unary constraints $x_j < n_j$ ( $x_j \leq n_j$) of $eg(\tau_{s,0})$ guarantee that each of the constraints in the guard $g'_{s,0}$ of the silent transition $\tau_{s,0}$ can be satisfied separately at some time that is equal or is later than $t_s$.
Then, in order that all the constraints could be satisfied simultaneously, it suffices to show that the minimum upon the time delays to the upper bound constraints of the clocks appearing in $g'_{s,0}$ is greater than the maximum upon the time delays to the lower bound constraints in $g'_{s,0}$ (the 'greater' should be replaced by 'greater or equal' in case both the maximum and minimum come from weak inequalities):
\begin{equation}
\min_{j} (n_j - x_j) > \max_{i} (m_i - x_i).
\end{equation}
But this condition is equivalent to the condition that $n_j - x_j > m_i - x_i$ at time $t_s$ for every $i, j$, which is exactly the conjunction of diagonal constraints
\begin{equation}
\label{eq:enabling_guard}
\bigwedge_{i \neq j} x_j - x_i < n_j - m_i
\end{equation}
of $eg(\tau_{s,0})$.

Thus, we know that the silent transition $\tau_{s,0}$ can be taken in the run $\rho$ at some time $t_{s,0}$ after a delay of $M = \max_{i} (m_i - x_i)$ from $t_s$ (this delay is not negative since we introduced the constraint $0 \leq x_s$) and before a delay of $N = \min_{j} (n_j - x_j)$. 

It remains to show that the transitions $\tau_{s+1}, \ldots, \tau_{s+p}$ on guards $g_{s+1}, \ldots, g_{s+p}$ of $\rho$ can be taken at the same dates $t_{s+1}, \ldots, t_{s+p}$ as the corresponding transitions on guards $g'_{s+1}, \ldots, g'_{s+p}$ are taken in $\rho'$. 
%

To be more specific, it suffices to prove that there exists $t_{s,0}$ with the following conditions:
\begin{enumerate}
\item $t_s \leq t_{s,0} \leq t_{s+1}$;
\item $g'_{s,0}$ is satisfied at $t_{s,0}$;
\item the constraints on $x_{s,0}$ are satisfied at $t_{s+1}, \ldots, t_{s+p}$, with $x_{s,0}$ reset at $t_{s,0}$.
\end{enumerate}

For condition 2. the constraints of $g'_{s,0}$ that should be satisfied at time $t_{s,0}$ are
\begin{equation}
\bigwedge_{i=1, \ldots, r} m_i < x_i(t_{s,0}) < n_i.
\end{equation}
Equivalently, at each time $t_{s+j}$, $j = 1, \ldots, p$:
\begin{equation}
\bigwedge_{i=1, \ldots, r}  m_i + t_{s+j} - t_{s,0} < x_i(t_{s+j}) < n_i + t_{s+j} - t_{s,0},
\end{equation}
or,
\begin{equation}
\label{eq:silent_tans_cnstr}
\bigwedge_{i=1, \ldots, r} m_i - x_i(t_{s+j}) + t_{s+j} < t_{s,0} < n_i - x_i(t_{s+j}) + t_{s+j}.
\end{equation}

For condition 3. the constraints on $x_{s,0}$ that should be satisfied at times $t_{s+1}, \ldots, t_{s+p}$ are $m_{s+j} < x_{s,0}(t_{s+j}) < n_{s+j}$ for $j = 1, \ldots, p$.
The constraint here at time $t_{s+1}$ is $0 \leq x_{s,0}(t_{s+1})$ possibly conjuncted with other constraints (for convenience we wrote all constraints as strict inequalities).
This is equivalent to
\begin{equation}
\bigwedge_{j = 1, \ldots, p} m_{s+j} < t_{s+j} - t_{s,0} < n_{s+j}
\end{equation}
or
\begin{equation}
\label{eq:reset_clk_cnstr}
\bigwedge_{j = 1, \ldots, p} -n_{s+j} + t_{s+j} < t_{s,0} < -m_{s+j} + t_{s+j}.
\end{equation}

We need to show that the constraints on $t_{s,0}$ of~(\ref{eq:silent_tans_cnstr}) and ~(\ref{eq:reset_clk_cnstr}) do not define an empty set.
This condition is equivalent to showing that the set $S_1$ of the above expressions to the left of $t_{s,0}$ is smaller than the set $S_2$ of the expressions to the right of $t_{s,0}$ (equivalently that the maximum of $S_1$ is smaller than the minimum of $S_2$), where
\begin{equation}
S_1 = \{ m_i - x_i(t_{s+j}) + t_{s+j} \, | \,  i = 1, \ldots, r, \, j = 1, \ldots, p \} \cup \{  -n_{s+j} + t_{s+j}  \, | \,  j = 1, \ldots, p \},
\end{equation}
and
\begin{equation}
S_2 = \{ n_i - x_i(t_{s+j}) + t_{s+j} \, | \,  i = 1, \ldots, r, \, j = 1, \ldots, p \} \cup \{  -m_{s+j} + t_{s+j}  \, | \,  j = 1, \ldots, p \}.
\end{equation}
There are two types of expressions in $S_1$ and two types of expressions in $S_2$, hence we need to check that the following $4$ cases are satisfied.
\subsubsection{Case 1: $m_i - x_i(t_{s+j}) + t_{s+j} < n_{i'} - x_{i'}(t_{s+j'}) + t_{s+j'}$.}
This inequality is equivalent to
\begin{equation}
m_i - x_i(t_{s,0}) + t_{s,0} < n_{i'} - x_{i'}(t_{s,0}) + t_{s,0},
\end{equation}
or to
\begin{equation}
m_i - x_i(t_{s,0}) < n_{i'} - x_{i'}(t_{s,0}).
\end{equation}
The latter is equivalent to
\begin{equation}
x_{i'}(t_s) - x_i(t_s) < n_{i'} - m_i,
\end{equation}
which is~(\ref{eq:enabling_guard}), the enabling guard $eg(\tau_{s,0})$ that is satisfied at time $t_s$ of the run $\rho'$. 
\subsubsection{Case 2: $m_i - x_i(t_{s+j}) + t_{s+j} < -m_{s+j'} + t_{s+j'}$.}
This inequality is equivalent to
\begin{equation}
m_i - x_i(t_{s+j'}) + t_{s+j'} < -m_{s+j'} + t_{s+j'},
\end{equation}
\begin{equation}
m_i - x_i(t_{s+j'}) < -m_{s+j'},
\end{equation}
\begin{equation}
 m_i + m_{s+j'}< x_i(t_{s+j'}).
\end{equation}
The last inequality is no other than one of the left inequalities of~(\ref{eq:updated_guard}), which are the updated future constraints in $A'$ of the reset clock $x_{s,0}$, and thus are given to be satisfied.
\subsubsection{Case 3: $ -n_{s+j'} + t_{s+j'} < n_i - x_i(t_{s+j}) + t_{s+j} $.}
This inequality is equivalent to
\begin{equation}
-n_{s+j'} + t_{s+j'} < n_i - x_i(t_{s+j'}) + t_{s+j'},
\end{equation}
\begin{equation}
-n_{s+j'} < n_i - x_i(t_{s+j'}),
\end{equation}
\begin{equation}
  x_i(t_{s+j'}) < n_i + n_{s+j'}.
\end{equation}
But the last inequality is one of the right inequalities of~(\ref{eq:updated_guard}), which are the updated future constraints in $A'$ of the reset clock $x_{s,0}$, and thus are given to be satisfied.
\subsubsection{Case 4: $ -n_{s+i} + t_{s+i} < -m_{s+j} + t_{s+j} $.}
This inequality is equivalent to
\begin{equation}
m_{s+j} - n_{s+i} < t_{s+j} - t_{s+i}.
\end{equation}
The inequality certainly holds when $i = j$.
When $i < j$ we can write this inequality with the clock $x_{s+i}$ that is reset at time $t_{s+i}$ in $A'$:
\begin{equation}
m_{s+j} - n_{s+i} < x_{s+i}(t_{s+j}).
\end{equation}
But the last inequality can be found in the first row of Table~\ref{table:updating_guards_2} which contains the synchronization constraints of the updated future constraints in $A'$ of the reset clock $x_{s,0}$.

Similarly, when $j < i$ we need to satisfy the inequality
\begin{equation}
 x_{s+j}( t_{s+i}) = t_{s+i} - t_{s+j} < n_{s+i} - m_{s+j},
\end{equation}
which can be found in the forth row of Table~\ref{table:updating_guards_2}.

We showed that the set of possible time values $t_{s,0}$ for the silent transition in $\rho$ is not empty, that is, there is a solution to the set of inequalities (\ref{eq:silent_tans_cnstr}) and (\ref{eq:reset_clk_cnstr}) in the indeterminate $t_{s,0}$ (again, the extension to weak inequalities is straight forward).

To complete the proof it remains to show that the solution for $t_{s,0}$ satisfies condition 1., that is that $t_s \leq t_{s,0} \leq t_{s+1}$.
Well, the left inequality $t_s \leq t_{s,0}$ comes from satisfying the inequality $m_i - x_i(t_{s+j}) + t_{s+j} \leq t_{s,0}$ of (\ref{eq:silent_tans_cnstr}) with $x_i = x_s$ and $m_i = m_s = 0$ (it refers to augmenting the silent transition guard with the constraint $0 \leq x_s$).
This inequality is equivalent to $0 -x_s(t_s) + t_s \leq t_{s,0}$ or $t_s \leq t_{s,0}$ since $x_s$ was reset at time $t_s$.

The right inequality comes from satisfying the inequality $t_{s,0} \leq -m_{s+1} + t_{s+1}$ of (\ref{eq:reset_clk_cnstr}) with $m_{s+1} \geq 0$, that is, $t_{s,0} \leq t_{s+1}$.
%
%
%



%
%
%
\subsection{Proof of Theorem~\ref{th:GOdet} [Determinization]}
The deterministic property of $D(A)$ follows from the fact that when merging $\alpha$-transitions into $\tau_{acc}$ and $\tau_{\neg acc}$ then the guard of $\tau_{\neg acc}$ is a conjunction of some guard with the negation of the guard of $\tau_{acc}$.
Hence, different runs will induce different time traces.

In general, by merging locations of $A$ in $D(A)$ we may only expand the language and conclude that $\LLL(A) \subseteq \LLL(D(A))$.
On the other hand, the new constraints introduced in $D(A)$ may restrict the language.
So, let us examine the new transformed constraints and show that they do not impose additional restrictions.
Suppose the guard of transition $\tau$ contains the constraint $x \sim n$ and that $y$ is reset on $\tau$.  
Then, at the time $t_0$ of $\tau$, the constraint $x(t_0) - y(t_0) \sim n$ holds.
But also at time $t_1 > t_0$, the constraint $x(t_1) - y(t_1) \sim n$ holds since $x$ and $y$ progress at the same rate.
Hence, for any run through $\tau$ in $A$ there exists a corresponding run in $D(A)$ with the same trace because the additional constraints of the form $x - y \sim n$ that are added to the future guards are satisfied automatically by all runs in $D(A)$ that satisfy the guard of $\tau$.
Thus, it remains $\LLL(A) \subseteq \LLL(D(A))$.

To show that the language of $D(A)$ does not contain accepting traces that are not in the language of $A$ it suffices to show that when a transition in a merged location of $D(A)$ is enabled then the corresponding original transition in $A$ is enabled.
But this is indeed the case since for each transition of $D(A)$ we first copy to its guard the transformed guard of the transition that leads to it, and this transformed guard contains all the history: the transformed guards of the path that leads to this transition.
That is, by induction one shows that since the record of paths of level $n$ are passed to paths of level $n+1$ then it holds for every level.  

\end{document}